# Influence of additives (CoO, CaO, $B_2O_3$) on thermal and dielectric properties of BaO-$Al_2O_3$-$SiO_2$ glass-ceramic sealant for OTM applications

A.I. Borhan[a]*, M. Gromada[a], G.G. Nedelcu[b], L. Leontie[b]

[a]*Institute of Power Engineering, Ceramic Department CEREL, Research Institute, 36-040 Boguchwała, Poland*
[b]*Faculty of Physics, Alexandru Ioan Cuza University of Iasi, 11 Carol I Boulevard, 700506 Iasi, Romania*
*Corresponding author: Tel/Fax: +48 17 87 11 700 / +48 17 87 11 277 (A.I. Borhan)
E-mail: adi.borhan@yahoo.com / borhan@cerel.pl

**Abstract**

The efficiency of glass ceramic sealants plays a crucial role in high temperature applications performance and durability, such as oxygen separation from air. In order to develop suitable sealants, operating at about 950°C, several glass-ceramics with different compositional additives and ratios in BaO–$Al_2O_3$–$SiO_2$ system have been prepared and investigated. The effects of additive oxides (CaO, CoO, $B_2O_3$) as network modifiers on the thermal and dielectric properties of BaO-$Al_2O_3$-$SiO_2$ glass-ceramic for further oxygen transport membrane (OTM) applications were evaluated and discussed. The introduction of additives has evident effects on the main crystalline phases and physical properties of the glass-ceramics. Among these systems, $B_2O_3$-rich BAS system (BABS) can be densified at the sintering temperature below 700°C through viscous sintering. Also, BABS shows the lowest dilatometric softening temperature at 675°C, which can be explained by a fast densification associated with a shrinkage of 25%. It has highest conductivity and resistivity, a high dielectric constant of 67.8 and a loss of 0.029 at 10MHz, which provide an attractive insulating feature for many technological applications. The electrical results obtained for BABS glass-ceramic are consistent with the typical requirements for an excellent sealant. However, considering all other properties investigated, only $B_2O_3$-rich BAS meets the requirements for use as a sealant for BSCF membrane. Low softening temperature of BABS opens more ways for adding additives to tailor its functional properties, which enable for using as sealant material for OTM applications.

**1. Introduction**

The development of glass-ceramic sealants for bonding oxygen membranes with other elements in device for oxygen production presents a significant challenge since they must meet very restrictive requirements. They must ensure at high temperature, reaching 950°C the mechanical strength, good bonding, chemical stability under operating environment. Gas tight sealants are vitally important for the performance, durability and safety operation of the installation for oxygen separation from air. In fact, the sealing of the glass-ceramic with the metallic and ceramic elements, which is necessary to obtain a hermetic joint, still remains a critical point in the fabrication of device for oxygen separation from air [1].

Among the most promising materials for sealing, glasses or dense glass-ceramics, which "glue" the components together by forming a rigid bond and provide hermetic sealing, have been studied. Glass-ceramics are considered advantageous for sealant applications because, in principle, meet most of the requirements of an ideal sealant [2] and [3]. Above the glass softening temperature, they soften and flow to form an intimate contact with surface of the metallic and ceramic components to be sealed and maintain the sealing performance. However, they harden and tend to detach from sealing surfaces or crack if their coefficient of thermal expansion (CTE) differs from those of the contacting components. Therefore, the CTE of the rigid sealing materials must be carefully tailored by optimizing their composition.

Barium aluminosilicates have suitable properties to be used as glass-sealant. Several compositions have been studied in the literature [4], [5], [6] and [7]. In the last decade, several innovative glasses have been also developed to identify the respective contribution and influence of each addition element. Many attempts were made to reduce the firing temperature by the addition of various low melting and low loss glasses [8] and [9]. Most importantly, the presence of $Al_2O_3$ in parent glass elevates the sintering temperate, so the application range of these glasses is limited [10]. Improving the mechanical property, decrease of sintering temperature and the production cost are becoming a hot topic. Moreover, it has been reported [11] that $B_2O_3$ decreased glass viscosity, delayed crystallization giving better wettability of glasses on the steel. $B_2O_3$ addition increases the fraction of non-bridging oxygen containing borate and silicate structural units. Therefore, the higher amount of non-bridging oxygen decreases network connectivity and thus softening temperature and melting point [12]. The gradual increase of CoO content in the glass matrix caused a decrease of structural compactness of the material, that influences the density of the glass ceramic [13]. The introduction of CaO in system is expected to increase the melting point, and the dielectric constant, but makes the vitrification difficult [14] and [15].

But, most investigations have taken into consideration the performance of glass-ceramic materials in SOFC applications [6], [12], [16], [17] and [18]. Therefore, it is imperative the development of glass-ceramic seal manufacturing technologies for application in device for oxygen separation from air. Obviously, connecting and sealing of the membrane systems at high temperatures is one of the major issues. It has been noted that the studies on high-temperature seals for mixed conducting membrane is rarely reported in literature. Gromada et al. [19] reported that best solution for use of BSCF membrane as oxygen transport membrane is in the shape of tube. Sealing of the materials is difficult because of the high pressure differential although the sealing area per unit area of membrane surface is small in a tubular design. It is vital to note that a large temperature gradient would exist over the separation and sealing sections of the apparatus. The use of glass-ceramic materials as a sealant has been employed due to mechanical, chemical, and electrical properties that can withstand high-temperature operating condition [20]. Various efforts to seal solid oxide ion conducting devices have been made with varying degrees of success. Silica, boron, and alumina-based glasses and glass-ceramics have been evaluated as a sealing materials for high temperature applications [3], [21] and [22]. Qi et al. reported that a rate for sealing the ceramic membranes of nearly 100% is possible using the glass-ceramic composite recipe if the correct sealing procedure, including seal paste preparation, is carefully followed [23]. Thursfield and Metcalfe used glass-ceramic seals that consist of boro-silicate glass for sealing the linear hollow-fiber membranes [24]. In another type of approach, Tan et al. [25] used silicone sealant for the oxygen production through $La_{0.6}Sr_{0.4}Co_{0.2}Fe_{0.8}O_{3-\delta}$ perovskite membranes. Although, some of seals helps to make perovskite oxygen membrane gastight at varying temperatures for oxygen permeation testing in laboratories, they cannot tolerate high temperatures and high pressures, or large difference of pressure at high temperature, which make them unpractical in the industrial applications.

In this paper, the effect of additive oxides (CaO, CoO, $B_2O_3$) as network modifiers on the thermal and dielectric properties of $BaO$-$Al_2O_3$-$SiO_2$ glass-ceramic for further OTM applications was studied; the thermal expansion coefficient and softening temperature of the glass-ceramics were tested and analysed.

**2. Experimental details**

The chemical compositions of the investigated glass-ceramics are listed in Table 1. The composition $25BaO$–$10Al_2O_3$–$65SiO_2$ (wt%) (BAS) was used as basic glass, and different amounts of CoO, CaO and $B_2O_3$ were introduced to replace $SiO_2$ and $Al_2O_3$ in the parent glass. The amount of $Al_2O_3$ has a significant effect on the thermal properties of the glass up to 10%. Exceeding this value leads to fast crystallisation of glass upon cooling. The content of CoO was limited at 7% because it affects the melting point and density of the glass material. Addition of CaO in samples increases the melting point and makes the vitrification difficult. The large amount of CaO in the samples exceeding 10% is detrimental for dielectric properties and melting point. The content of the $B_2O_3$ was limited at 30 wt%, since it is detrimental to the dielectric properties, as reported in the literature [26].

Table 1. The effect of CoO, CaO, $B_2O_3$ ratios on the melting temperature of BAS glass.

| Sealant | Kind of glass (wt%) | Heat treatment | Melting or not (950°C, 1h) |
|---|---|---|---|
| BAS | 25BaO, 10 $Al_2O_3$, 65 $SiO_2$ | 1100°C/7h +1500°C/7h | No |
| BACoS | 25BaO, 8 $Al_2O_3$, 10 CoO, 60 $SiO_2$ | 1100°C/7h +1500°C/7h | No |
| BACaS | 25BaO, 10 $Al_2O_3$, 5CaO, 60 $SiO_2$ | 1100°C/7h +1500°C/7h | No |
| BABS | 25BaO, 10 $Al_2O_3$, 30$B_2O_3$, 35$SiO_2$ | 1100°C/7h | Yes |

Therefore, the following statement was added in the experimental procedure. Analytical reagent graded barium oxide BaO, aluminium oxide $Al_2O_3$, silica $SiO_2$, cobalt oxide CoO, calcium oxide CaO, and boron $B_2O_3$ as purchased from Sigma-Aldrich without further purification steps (purity > 99.9%), were used to prepare the glass-ceramic samples. In this context, the Co-dopped BAS glass is designated as BACoS, Ca-dopped glass is named as BACaS, and B-rich glass is designated as BABS. The starting raw materials were mixed for 60 min in a ball mill, and then transferred to an alumina crucible and molten in the temperature range of 1100-1500°C with a dwell time of 7h. The samples obtained at 1100°C are shown in Fig. 1. It is clearly observed that only BABS was obtained in the form of glass at 1100°C. The other samples were further sintered at 1500°C for 7 h.

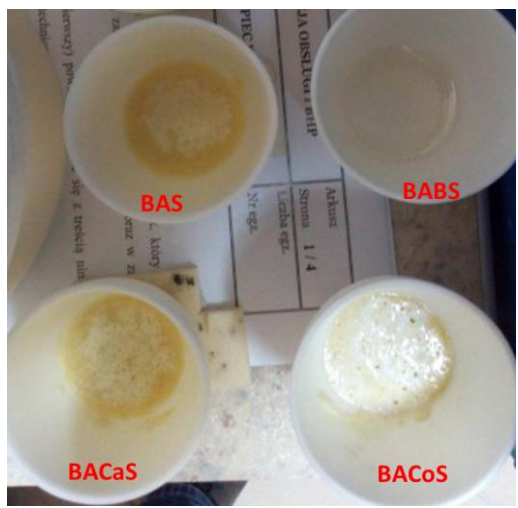

Fig.1. The samples obtained at first stage of sintering, 1100°C

The batch size in each run was 30 g resulting in 15-20 g of final product. The resulting melts were slightly cooled up to 800°C with a slow cooling of 100°C/h, followed by free cooling to room temperature. The glasses obtained were crushed and milled in dry conditions in a rotating mill using agate containers and agate balls for 2 h to obtain a fine glass powder. The bulk glasses were machined into bar specimen (7 mm in diameter, and 10 mm in length) for CTE measurement, and were milled into powders for the heating microscopy, granulometry test, X-ray diffraction, and wetting test. Softening behaviour and wettability of glasses were investigated by heating microscope (Carl Zeiss, Jena model MH0-2) from room temperature to 1100°C with a heating rate of 10° min$^{-1}$. The coefficient of thermal expansion in the range of temperature from 25 to 950°C was determined using high temperature dilatometer from BÄHR-Gerätebau GmbH at a heating rate of 10° min$^{-1}$. The glass-ceramic powders were then subjected to XRD phase analysis using X'PertPROX diffractometer with CuK$_\alpha$ radiation ($\lambda$=1.5406 Å), for 2$\theta$ ranging between 10 and 80°, at a scanning speed of 0.02°/s, to confirm their amorphous nature. Particle size distributions were measured from wet dispersions using large (Malvern Hydro 2000 MU) volume sample dispersion unit available for the Mastersizer 2000 granulometer. Sealing capability of fabricated glasses was assessed by means of a very simple way. For this purpose, 50 mg of glass powder were sandwiched in an assembly made of stainless steel and tubular membrane. The assembly was placed in a furnace, and heated up to 950 °C for 1 h, and then cooled down to room temperature.

Dependence of dielectric permittivity and dielectric losses as a function on applied field frequency was studied by using an Agilent 4292-1 device, in the range of 40 Hz–10 MHz. The pellets were pressed into a cylindrical disk at 300 kPa/cm$^2$, without subsequent calcination, and were inserted between two flat electrodes.

## 3. Results and discussion

Fig. 2 shows the XRD patterns of samples BAS, BACoS, BACaS and BABS heat-treated under different conditions.

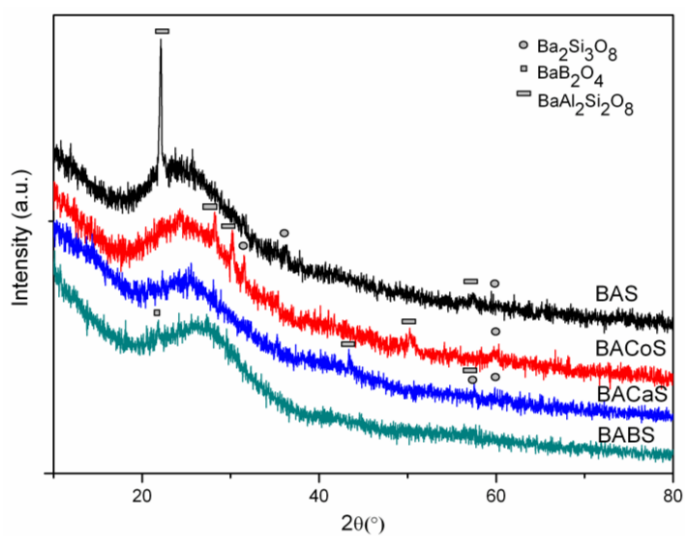

Fig.2. XRD patterns of BAS glass-ceramics containing different additives (CoO, CaO and $B_2O_3$)

The broaden peak was found between $2\theta = 18\text{-}30°$ corresponding to the amorphous glass in the microstructure. It can be observed that all samples were basically composed of glass phases and there were little differences between BACoS, BACaS and BABS glass powders. It seems that BACoS, BACaS and BABS possess more glass nature, due to less crystallite content. These observations are based on the presence of a set of high and low-intensity XRD peak, distributes in glass matrix, symbolized by a broad and diffuse peak. The samples start crystallization to over 800°C, and the crystallization capacity is improved with the increase in temperatures and holding time [15]. It can be observed an intense diffraction peak at 2θ=22°, corresponding to hexacelsian $BaAl_2Si_2O_8$ (JCPDS No. 01-088-1051) phase in case of BAS sample, coexisting with $Ba_2Si_3O_8$ phase (JCPDS No. 12–0694). Introduction of glass modifiers CoO, CaO and $B_2O_3$ leads to decrease in the intensity of these peaks, which corresponds to a much smaller presence of crystalline hexacelsian phase. The XRD pattern of the glass BABS revealed that no $BaAl_2Si_2O_8$ or $Ba_2Si_3O_8$ phases but $BaB_2O_4$ (JCPDS card No. 80-1489) was detected. The $BaAl_2Si_2O_8$ was not detected in BABS glass because of boron possibly associated in the hexacelsian crystal structure during the process of crystallization [27].

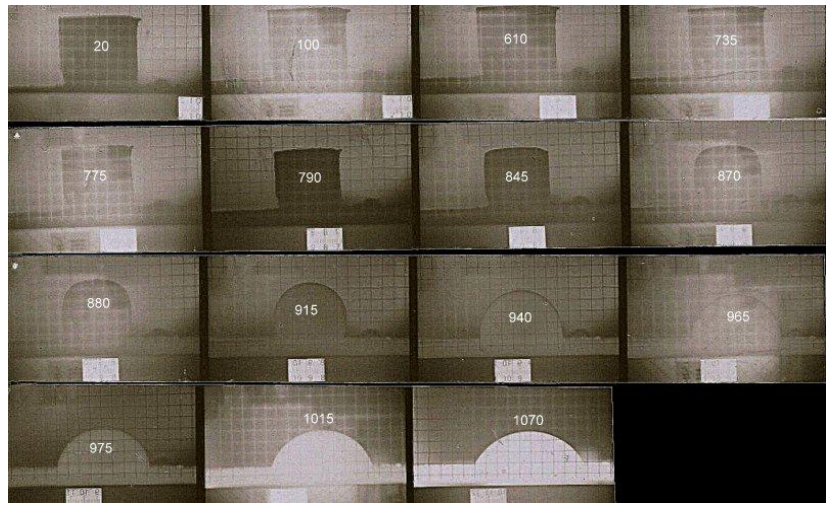

Fig.3. Shape change of BABS glass pellet as a function of temperature

Micrographs in Fig. 3 are representative pictures of bonding characteristics and softening behaviour of the BABS glass pellets with increasing temperature. The geometric shapes of the glasses began to change with shrinkage initiation, $T_{SI}$, when the pellet started to contract, at 735°C. Maximum shrinkage of the glass pellet takes place at the temperature of maximum shrinkage, $T_{MS}$=845°C before it starts to soften. Softening point, $T_S$=870°C, is the temperature at which the first signs of softening are observed by rounding of the pellet. The next important step is half ball temperature, $T_{HB}$=965°C, where the section view of samples became like a semicircle. At the final stage, pellet flowed on the substrate at flowing temperature, $T_F$=1070°C. Note that wetting angle decreased with temperature and better wettability was observed for BABS glass at about 1070°C with a contact angle near 50°. It can be observed that around OTM sealing temperatures (≈950°C) shapes of the pellets became rounded but the measured wetting angle is > 90°. Generally, higher temperatures for sealing process can be helpful to attain better bonding [28], but can affect the metallic part of installation [29]. Dilatometric results for all glasses, which indicate the shrinkage behaviour of the glass sealants versus temperature, are shown in Fig. 4.

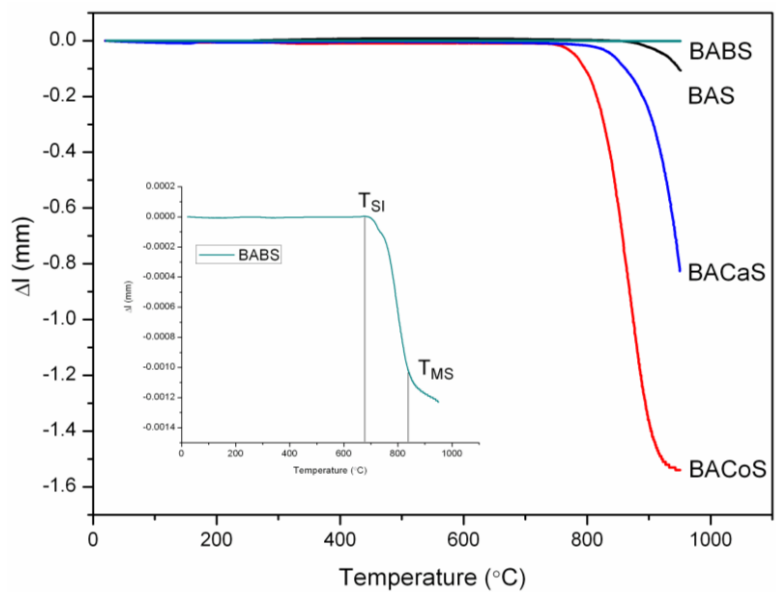

Fig.4. Sintering shrinkage curves of BAS samples containing various additives (insert is shown shrinkage curve of BABS)

All glasses in this study give rise to glass–ceramics after sintering. It can be clearly seen that for all the glasses, the maximum amount of shrinkage is within the range of 10–25%, and the shrinkage starts from 675°C, and attains their highest level within 860°C. BACoS has a low dilatometric softening temperature ($T_s$) at 770°C, which produced a rapid densification associated with greater than 20% shrinkage. Apparently, the densification process is not affected by crystallization of hexacelsian $BaAl_2Si_2O_8$ phase. Instead, the BAS glass consistently shrinks within 880–950°C and a very slight decrease in shrinkage has been observed at the sealing temperature (950°C), due to the existence of hexacelsian phase, suggesting very little flow of the glass at this temperature (see Fig. 6). Therefore, this glass may be suitable for sealing at relatively higher temperature (>950°C). Incorporation of additives lowers the temperature where shrinkage occurs, but even so, the temperature is relatively high. For the BACoS and BACaS glasses, the maximum shrinkage is observed at 920°C and 950°C. After that, the materials entered in the expansion process before melting completely. BABS shows the lowest dilatometric softening temperature at 675°C, which can be explained by a very fast densification associated with a shrinkage of 25%. The level of the sintering shrinkage is correlated to the locations of the glass softening point and the crystallization point. The beginning of shrinkage at a higher temperature, for un-dopped BAS, Co-dopped BAS and Ca-dopped BAS glass-ceramics, to over 770°C is probably due to its higher transition and softening temperatures, respectively [30]. It can be observed that none of the curve do not show reflection point or shoulder, which means that the densification process not proceeded simultaneously with crystallization [15]. This is confirmed by XRD analysis which showed that all materials are amorphous with some trivial crystalline phases.

As it can be seen in the case of BAS, the shrinkage was slowed down or stopped relatively quickly, which means that a large number of crystallites, due to the hexacelsian phase, inhibited particle rearrangement during the densification process. Therefore, it needs a much higher sintering temperature to cause more shrinkage [30]. It is generally considered that powders with a smaller particle size experienced a significant density decrease at the higher temperatures. The average (d(0.5)) particle size of the glass-ceramic powders used in the present investigation lies between 13.4–51.2 μm and their particle size distribution is presented in Fig. 5. The particle size distributions of all glass-ceramics can be grouped as normal - Gaussian particle size distribution.

The BABS had a higher particle size and small specific surface area, which means a decrease in density with $B_2O_3$ introduction in BAS system. For the rest of glass-ceramics, the probability of surface crystallization and formation of refractory crystalline phases on the surface, as revealed XRD, is enhanced by the smaller particle size and the increase of the nucleation sites. The specific surface area of BABS showed a significant reduction compared to other compounds, due to the coalescence of glass particles before crystallization [31]. All glass-ceramics can bond and wet substrate if the temperature is high enough.

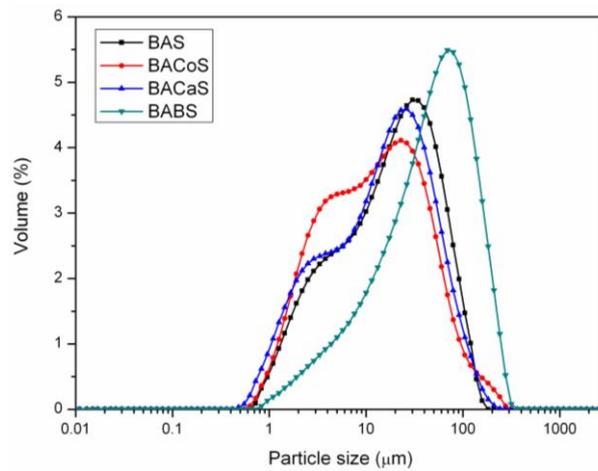

Fig.5. The particle size distribution of BAS glass-ceramics containing different additives

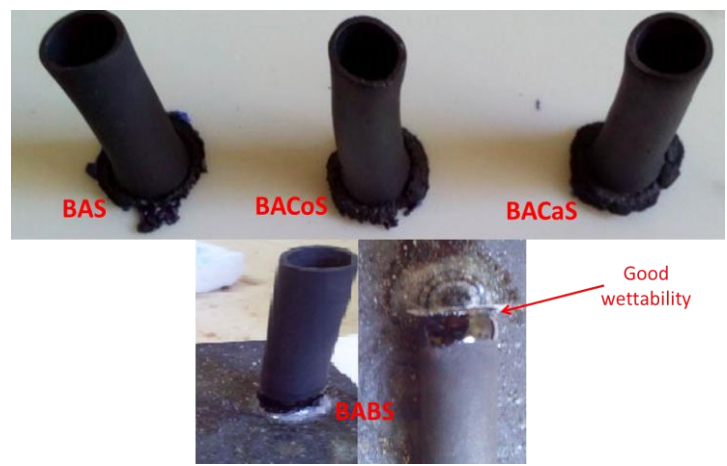

Fig.6. Picture of sealing capability test of glass-ceramics carried out at 950°C for 1h

Fig. 6 shows the picture of sealing test carried out at 950°C for one hour. The BAS, BACoS, BACaS glass-ceramics have a sealing temperature higher than 1000°C. This reduces to 950°C when the $B_2O_3$ oxide is introduced in system and content of $SiO_2$ decreases. It can be seen that the BAS, BACoS, and BACaS glass-ceramics are still powder at 950°C, indicating a poor wettability at this temperature. Instead, BABS glass-ceramic shows a very good sealing capability at 950°C, indicating this material as sealant for sealing the ceramic membranes.

Table 2 summarizes the effect of CoO, CaO and $B_2O_3$ additives on the thermal properties of BAS glass, obtained from dilatometric measurements.

Table 2 summarizes the effect of CoO CaO, and $B_2O_3$ additives on the thermal properties of BAS glass, obtained from dilatometric measurement.

| Glass | Softening temperature (°C) | Melting temperature (°C) | CTE ($\times 10^{-6}$°C$^{-1}$) | | | Observation |
|---|---|---|---|---|---|---|
| | | | 25°C | 250°C | 950°C | |
| BAS | 1210 | 1470 | 2.67 | 6.64 | -5.54 | Not wetting support |
| BACoS | 1460 | >1500 | 1.37 | 3.03 | -188.95 | Not wetting support |
| BACaS | 1100 | 1320 | 0.79 | 2.77 | -123.06 | Not wetting support |
| BABS | 675 | 950 | - | - | -181.78 | Wetting support, good sealing |

According to theories, the melting temperature of glass-ceramics would be lowered with introduction of $B_2O_3$ in glass matrix, but only in the company of another additive [32]. Our results showed that the melting temperature of BAS glass can be greatly lowered with 520°C by adding 30 wt% $B_2O_3$ alone. The results also showed that glass compositions with 5 wt% CaO and 7 wt% CoO could not be molten without slag. The melting point for these samples was at 1320°C and > 1500°C, respectively. These glass–ceramics may be considered as refractory if they can be used at temperatures above 1200°C. But for refractory glasses coefficients of thermal expansion (CTEs) are low to moderate ($(25–45)\times 10^{-7}$ C$^{-1}$, from 25 to 1000°C) [33]. In contrast, for our materials was determined an extremely low coefficient of thermal expansion, which slightly increases between 25°C and 250°C. It can be noted that ceramic BACaS with close to zero thermal expansion ($0.19 \times 10^{-6}$ $C^{-1}$) at 25°C was prepared by melt reaction. The decrease in CTE with introduction of new cations in matrix is believed to be associated with the change in the nature of bonding in the structural network, and it could thus be concluded that the CoO, CaO, and $B_2O_3$-added glasses sustain a high structural stability with the change in time [34]. In the literature, barium aluminosilicate crystals are known as having a great CTE ranging between 8.1-11.8 $\times 10^{-6}$ $K^{-1}$ [35] and [36]. Taking into account that BAS glass-ceramic is mainly composed of $BaSiO_3$ crystals and $BaAl_2Si_2O_8$, it is obvious that the small value of the CTE at 25°C is caused by the formation of these hexacelsian crystals (CTE of $BaAl_2Si_2O_8$ is around $2\text{-}3 \times 10^{-6}$ $K^{-1}$, according with literature) [21]. On the other hand, as the amount of glass modifiers (CoO, CaO, $B_2O_3$) increases in the glass by substituting equivalent amount of glass formers ($SiO_2$ and $Al_2O_3$), the number of non-bridging oxygen also increases within the glass matrix which, in turn, causes a lowering in the softening and melting temperatures [37].

On the other hand, the great surprise lies in the coefficient of thermal expansion values at 950°C. BACoS, BACaS and BABS glass-ceramics can be classified as giant NTE materials, presenting giant negative thermal expansion, i.e. $-188 \times 10^{-6} C^{-1}$, $-123 \times 10^{-6} C^{-1}$, and $-181,78 \times 10^{-6} C^{-1}$ at 950°C. BABS glass-ceramic shows a huge negative thermal expansion between 25°C and 950°C. In general, the thermal expansion behaviour of a glass–ceramics mainly depends on the thermal expansion of the crystalline phases, precipitated from the glass [38]. The thermal expansion of a glass is also related to the vibrations of the atoms in the material as a result of the change in the thermal energy. In solid glasses, the vibrations are restricted by the strong metal–oxygen bonds. The introduction of fillers in parent glass leads to drastic changes in the thermal expansion behaviour, even when the glass structure remains the same due to the formation of solid solutions composed of $Ba_2Si_3O_8$ and $BaAl_2Si_2O_8$. These solid solutions are assumed to possess CTEs which lie between the values of the pure compounds, and hence the CTEs should be tunable by changing the Al/Co, Al/Ca and Al/B ratios. But, in case of BABS, these crystalline phases were not found, which means that NTE may be due to other factors, such as Schottky, elastic, tunnelling and electronic effects [39]. Probably, a reason for the negative thermal expansion found in this glass-ceramics might be the strong displacement effects between aluminium and oxygen connecting two layers of the glass structure, i.e. Si/Al tetrahedral deformation [40]. Furthermore, the addition of boron to silica glass causes weakening of the structure because of the oxygen bridges between $Si^{4+}$ ions can be broken. However, these materials require more thorough research in order to fully characterize and perhaps to develop new giant NTE materials to run on a broader range of temperatures and thus in new applications. For many applications, NTE materials that have high electrical or thermal conductivity would be desirable [41].

Variation of real ($\varepsilon'$) of dielectric constant and dielectric loss in the frequency range 40Hz-12MHz for all glass-ceramics is shown in Fig. 6 (a-b). It has been observed that all glass-ceramics exhibit a normal dielectric dispersion, due to Maxwell-Wagner type interfacial polarization in accordance with Koop's phenomenological theory [42] and [43], that can be attributed to the decrease of dielectric constants with increasing frequency. This theory says that the conductivity of grain boundaries contributes more to the dielectric constant at lower frequencies. The dielectric properties are due to combined intrinsic and extrinsic factors: lattice vibration modes, porosity, chemical homogeneity, oxygen vacancies, grain size [44]. Domination of any of these factors varies with the sample's composition and sintering temperature [45].

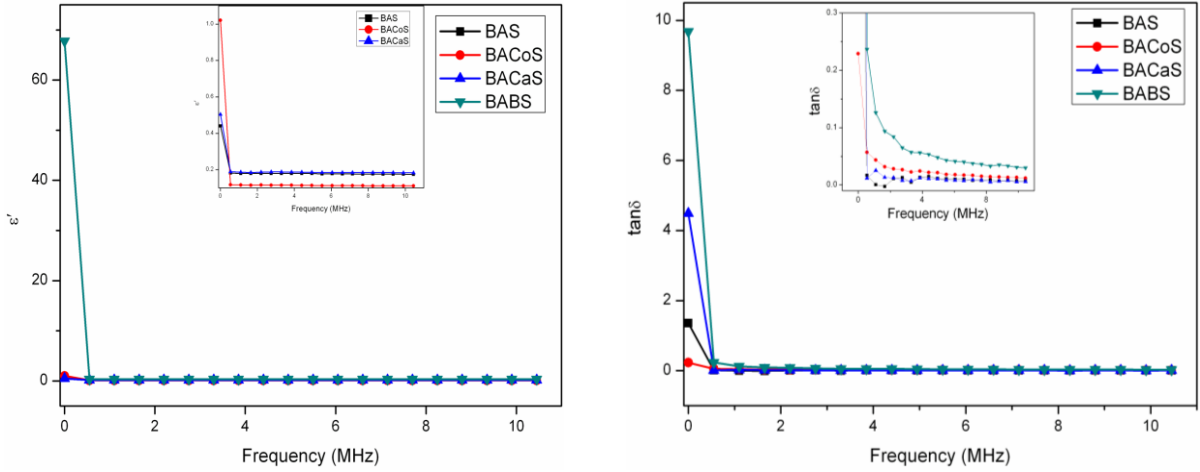

Fig.7. Variation of real part (ε') of dielectric constant and dielectric loss in the frequency range 40Hz-12MHz for all glass-ceramics

It can be observed that dielectric constants and losses first decrease with increasing frequency and then reach almost constant values at high frequencies. According to the Maxwell-Wagner model, this variation occurs because the hopping between different metal ions cannot follow the alternating field because of the predominance of species such as lattice defects, oxygen vacancies, grain size/boundary, secondary phases [46].

Table 3. Dielectric properties, particle size and specific surface area of investigated glass–ceramics.

| Glass | ε' | tanδ | σ (S/m) | Φ (Ωm) | Particle size (μm) | Specific surface area (m$^2$/g) |
|---|---|---|---|---|---|---|
| BAS | 0.44 | 0.007 | 3.6×10$^{-9}$ | 2.76×10$^9$ | 19.9 | 0.768 |
| BACoS | 1.02 | 0.012 | 9.7×10$^{-9}$ | 1.02×10$^9$ | 13.4 | 0.938 |
| BACaS | 0.50 | 0.006 | 4.6×10$^{-9}$ | 2.15×10$^9$ | 16.6 | 0.964 |
| BABS | 67.8 | 0.029 | 1.4×10$^{-6}$ | 6.7×10$^6$ | 51.2 | 0.339 |

Table 3 shows the dielectric properties and discussed earlier particle size and specific surface area of investigated glass–ceramics. The dielectric constant and loss are also markedly changed depending on the type of fillers. The CaO-rich glass-ceramic, BACaS, is expected to have a higher dielectric constant, due to the polarizability of the Ca$^{2+}$ (3.16 Å$^3$) ions, which are much higher than those of Co$^{2+}$ (1.66 Å$^3$) ions, B$^{3+}$ (0.05 Å$^3$) ions and Si$^{4+}$ (0.87 Å$^3$) ions [47]. Low dielectric constants of the BAS, BACoS and BACaS glass–ceramics, i.e. 0.44, 1.02 and 0.5, respectively, provide an attractive feature for minimizing cross talk and increasing signal transmission speeds. BAS glass–ceramic has the lowest dielectric constant, while its dielectric loss is also very low (tan $\delta$ = 0.007). According to the literature, $SiO_2$-rich glass–ceramics presents the lowest dielectric constant values because $SiO_2$ possess the lowest dielectric constant among the ceramics. Moreover, BAS has an extremely low tan $\delta$ among the glasses prepared in the high frequencies due to its rigid bonds [14]. BABS glass-ceramic possesses the highest dielectric constant ($\varepsilon$ = 68), while its dielectric loss is also the highest (tan $\delta$ = 0.029). $SiO_2$ and $B_2O_3$, which are glass formers, only generates limited elastic shift under electric field [48]. In addition, silica glass has the lowest loss among the glasses in the high frequencies region due to its rigid bonds. Therefore, it would be expected that BABS will show the lowest value of the dielectric losses. However, the introduction of $B_2O_3$ in the system to lower the melting temperature, causes the breaking of rigid bonds of $SiO_2$, leading to higher losses [14]. Particularly, BAS, BACoS and BACaS glass–ceramics contain hexacelsian phase which has a low dielectric constant [49]. As can be seen in Fig. 6a (insert), BACoS has the highest dielectric constant among these three samples, due to the fact that the CoO content will increase the polarization of ionic shift [50]. On the other hand, the high value of dielectric constant in case of $B_2O_3$-rich BAS system cannot be explained based on any secondary phase, but may be related with particle size. In this regard, glass-ceramic with higher particle size (lower densities) have higher dielectric constant and low dielectric loss. The introduction of $Co^{2+}$, $Ca^{2+}$ and $B^{3+}$ cations in BAS system could generate oxygen vacancies in the glass structure, and thus, the dielectric behaviour is not intrinsic in character and may be induced by structure defects. Hence, BABS glass-ceramics exhibit space charge polarization, which is thought to be arising from the difference between the grain boundaries conductivity of various phases (see Fig. 2). This causes a build-up of charge carriers at the interface, which corresponds to a high value of dielectric constant. The charge carriers accumulation at the grain boundaries is responsible for higher values of dielectric constant at low frequencies [51]. At high frequency, ε' results from the grains which have a small dielectric constant [52].

In agreement with results, BABS glass-ceramic with high dielectric constant and low loss with frequency could be considered as a possible candidate for many technological applications [53]. The frequency dependent electrical conductivity for investigated glass-ceramics measured at room temperature is shown in Fig. 7. The conductivity pattern can be split into two parts: (i) at lower frequencies (below 500 KHz) no dispersion of conductivity is observed up to that frequency; (ii) at frequencies above 500 KHz high dispersion of conductivity is observed.

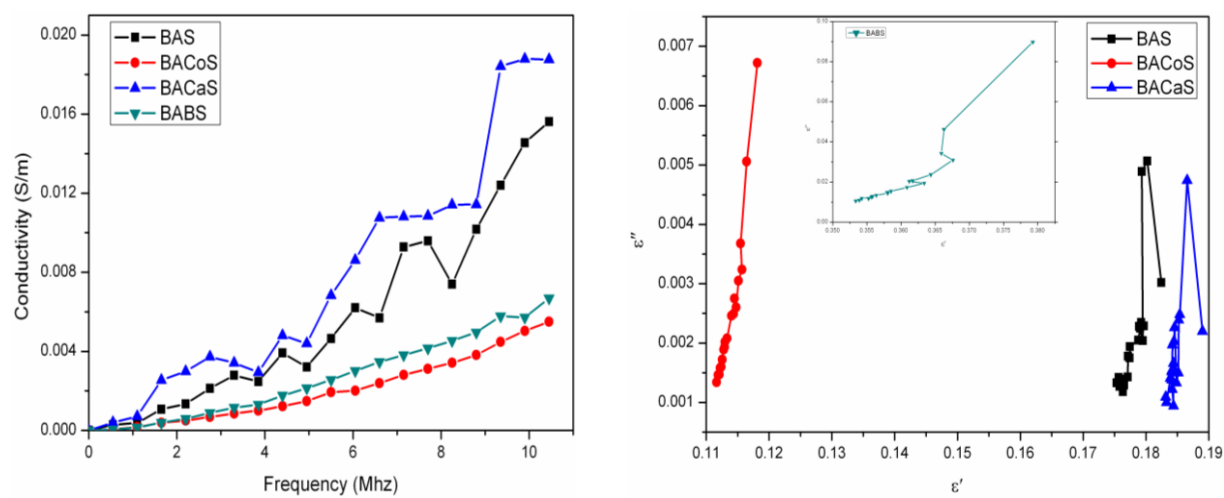

Fig8. (a) The frequency dependent electrical conductivity for investigated glass-ceramics measured at room temperature; (b) Cole–Cole plots of prepared glass-ceramics in the frequency range from 40Hz to 12MHz.

The conductivity increasing with frequency in glass-ceramics is generally due to hopping of charge carriers between different localized states [51]. In the present case, the observed increase in conductivity with frequency can be attributed to the increased hopping rate with increasing frequency. It is also evident that the frequency dependent ac electrical conductivity is strongly influenced by the filler added in the glass-ceramics structure.

The rise of ac conductivity function of fillers added, and in consequences with the decrease of the melting temperature (Table 2), indicates that the sintering process gives origin to structural modifications, which leads to an increase in the amount of particle charge carriers. The ac conductivity of BABS sintered at 1100 °C is higher than that of glass-ceramics sintered at 1500 °C indicating the existence of a higher number of dipoles that can follow the external electric field. According to the generally accepted theories of the constitution of glass, the basic structural elements of silica glass are $SiO_4$ tetrahedra, which are randomly oriented and do not possess symmetry. In binary alumina-silica glasses the ions added are assumed to be located at interstices between these tetrahedra. Since the bonding forces holding these ions in place are relatively weak, they have a high degree of mobility, which results in high conductivities and loss tangent in the glass-ceramic. From Table 3, it is evident that $B_2O_3$-rich glass has the higher dissipation factor at high frequency. These results can be explained by the assumption that the addition of $B^{3+}$ ions to silica glass causes some of the oxygen bridges between $Si^{4+}$ ions to be broken, and thus the glass structure is weakened. This means that the ions are able to move about with relative ease, and in consequence the conductivities and dissipation factor of such glass-ceramic are high.

Electrical resistivity (Table 3) of the BAS, BACoS and BACaS glass–ceramics was found to be slightly decreased, probably due to movement of mobile ions in structure [54]. Moreover, if silica is exchanged for $B_2O_3$ on a molar basis, it strongly decreases the electrical resistivity. This behaviour could be expected because boron oxide does not add additional mobile ions to the glass that can transport electric current. To the authors' knowledge, the influence of the interaction between boron oxide and alumina on the electrical resistivity of glass-ceramics has never been investigated. In our opinion the number of non-bridging oxygen sites is reduced significantly but at the same time, $Al_2O_3$ can „hinder" boron to interact with mobile modifying oxides. However, this is just a speculation, requiring extensive research. In conclusion, availability of charged particles increases the conductivity whereas the formation of space charge increases the dielectric constant [55].

Fig. 8 shows the Cole–Cole plots of prepared glass-ceramics in the frequency range from 40Hz to 12MHz. Complex dielectric constant plot (Cole–Cole plot) is using to distinguish between the contribution from grain boundary and grain itself [55]. For all glass-ceramic materials, the measurements confirm a capacitor effect dominating at low and mid-range frequencies. Nevertheless, an ideal capacitor should appear as a vertical straight line in the Cole–Cole plot. Note that BACoS seems to act as an ideal capacitor throughout the frequency range. Corroborating this result with the negative coefficient of thermal expansion indicates that BACoS can be used in many practical applications. The tilts of the lines indicate a small contribution of the capacitor losses in the dielectric. Moreover, it can be observed that only a single semicircle is present for BAS, BACaS and BABS samples. This suggests that the contribution of the grain boundary is predominant, while the contribution from the grain is not observed. However, the grain boundary contribution cannot be separated from the grain contribution in the electrical measurements [52] and [56]. From the variation of dielectric constant is clear that the values of grain boundary resistivity is almost constant for undoped BAS, Co-doped BAS and Ca-doped BAS, and then a decrease in the resistivity is produced. It can be noted that the grain boundary resistivity of examined samples is directly proportional to the polarizability.

The electrical results obtained for glass-ceramics are consistent with the typical requirements for an excellent sealant. However, considering all other properties investigated, only $B_2O_3$-rich BAS meets the requirements for use as a sealant for BSCF membrane. Starting from this material, various compositions have been already formulated to optimize thermal properties so as to achieve a perfect contact with the membrane BSCF [19].

**Conclusions**

Glass–ceramic materials within the system of $BaO–Al_2O_3– SiO_2$ (BAS) with various CoO, CaO and $B_2O_3$ additives and ratios have been studied as potential oxygen transport membrane sealants. The thermal and physical properties of all these glass-ceramics can be tailored by proper selection of composition particularly, by taking suitable amount of metal oxides.

The BAS, BACoS, BACaS glass-ceramics have a melting temperature higher than 1000°C. It has been observed that a relatively high boron $B_2O_3$ containing composition (BABS) could lead to a glass-ceramic having low softening and melting point. Instead, $B_2O_3$-rich BAS glass-ceramic shows a very good sealing capability at 950°C, indicating this material as sealant for sealing the ceramic membranes.

The maximum amount of shrinkage for all the glasses is within the range of 10–25%, and the shrinkage starts from 730°C and attains their highest level within 880°C. BABS was found to be most satisfactory in terms of flow and shrinkage behaviour.

The biggest problem lies in the coefficient of thermal expansion values at 950°C. All glass-ceramics can be classified as giant NTE material. Therefore, it is necessary to continue with the optimization of BABS composition, in order to obtain a coefficient of thermal expansion as close to the perovskite membrane for a perfect contact.

It can be concluded that the sintering temperature and additives introduction strongly influenced the dielectric properties and thermal properties of the BAS glass-ceramic. The results of dielectric constant were decreased with increasing sintering temperature up to 1500°C for BAS, BACoS and BACaS. It was also noticed that the dropped value of dielectric constant with higher sintered density at 1500 °C is due to the presence of crystalline phases during the sintering process. The dielectric constant and dielectric loss of BABS are higher than that of glass-ceramics sintered at 1500 °C, and can be correlated with the higher particle size. In agreement with results, BABS glass-ceramic with high dielectric constant and low loss with frequency could be considered as a possible candidate for many technological applications.

**Acknowledgements**

The research leading to these results has received funding from the People Programme (Marie Curie Actions) of the European Union's Seventh Framework Programme FP7/2007-2013/ under REA grant agreement No. PITN-GA-2013- 606878.

L. Leontie kindly acknowledge financial support from the ANCS (National Authority for Scientific Research), Ministry of Economy, Trade and Business Environment, through the National Program Capacities, Project No. 257/28.09.2010 (CERNESIM).